# Layer-controlled Ferromagnetism in Atomically Thin CrSiTe$_3$ Flakes

Cheng Zhang[+,1,4], Le Wang[+,2,3], Yue Gu[5], Xi Zhang[6], Liang-Long Huang[2,3], Ying Fu[2,3], Cai Liu[2,3], Junhao Lin[4], Xiaolong Zou[5], Huimin Su[*,2,3], Jia-Wei Mei[*,2,3,4], Jun-Feng Dai[*,2,3,7]

1. School of Physics, Harbin Institute of Technology, Harbin, 150001, China
2. Shenzhen Institute for Quantum Science and Engineering, Southern University of Science and Technology, Shenzhen, 518055, China
3. International Quantum Academy (SIQA), and Shenzhen Branch, Hefei National Laboratory, Futian District, Shenzhen, China
4. Department of Physics, Southern University of Science and Technology, Shenzhen, 518055, China
5. Shenzhen Geim Graphene Center, Tsinghua-Berkeley Shenzhen Institute & Tsinghua Shenzhen International Graduate School, Tsinghua University, Shenzhen 518055, China
6. Shannxi Institute of Flexible Electronics, Northwestern Polytechnical University, Xi'an 710072, China
7. Shenzhen Key Laboratory of Quantum Science and Engineering, Shenzhen 518055, China

[+] The authors contribute to this work equally
[*] Corresponding authors:
daijf@sustech.edu.cn; meijw@sustech.edu.cn; suhm@sustech.edu.cn

**Abstract:**

**The research on two-dimensional (2D) van der Waals (vdW) ferromagnets has promoted the development of ultrahigh-density and nanoscale data storage. However, intrinsic ferromagnetism in layered magnets is always subject to many factors, such as stacking orders, interlayer couplings, and the number of layers. Here, we report a magnetic transition from soft to hard ferromagnetic behaviors as the thickness of CrSiTe$_3$ flakes decreases down to several nanometers. Phenomenally, in contrast to the negligible hysteresis loop in the bulk counterparts, atomically thin CrSiTe$_3$ shows a rectangular loop with finite magnetization and coercivity as thickness decreases down to ~8 nm, indicative of a single-domain and out-of-plane ferromagnetic order. We find that the stray field is weakened with decreasing thickness, which suppresses the formation of the domain wall. In addition, thickness-dependent ferromagnetic properties also reveal a crossover from 3 dimensional to 2 dimensional Ising ferromagnets at a ~7 nm thickness of CrSiTe$_3$, accompanied by a drop of the Curie temperature from 33 K for bulk to ~17 K for 4 nm sample.**

## I Introduction

Layered ferromagnets coupled by van der Waals (vdW) force have recently gained considerable interest. Since weak interlayer interactions facilitate their mechanical exfoliation to monolayer or few-layer two-dimensional (2D) forms, they not only provide an ideal platform to study the fundamental magnetic interaction in 2D limit [1-6], but also offer great potential applications in the field of nanoscale spintronics and flexible electronics. However, as the number of layers of 2D magnets decreases, the intrinsic magnetism exhibits many peculiar characteristics, in contrast to the bulk counterparts. For example, layer-dependent ferromagnetic-to-antiferromagnetic transition in layered ferromagnets [2, 7], interlayer interaction induced change in magnetic order [8-10], and suppression of magnetic ordering in atomically thin layered antiferromagnets [11, 12]. For practical applications, it is necessary to search for much more robust 2D ferromagnets, which could remain its ferromagnetism down to atomically thin thickness.

$CrSiTe_3$ is an Ising-type ferromagnetic insulator with a Curie temperature ($T_c$) of 32.8 K[13-15]. It crystallizes in the $R\bar{3}$ space group, with the monolayers stacked in an ABCABC sequence (Fig. 1a). Within each layer, Cr atoms located at the center are closely sandwiched by Te atoms to form a 2D honeycomb network. The magnetic properties are mainly related to the magnetic moment of the Cr atom, which is around ~2.7 $\mu B$ [16]. However, the absence of remanent magnetization at zero external magnetic field makes it exhibit a soft ferromagnetic property, which limits its practical application on spintronic devices. Recently, pressure-enhanced ferromagnetism in layered $CrSiTe_3$ [17] and $CrGeTe_3$ [18] have been reported. However, the intrinsic ferromagnetism in atomically thin $CrSiTe_3$ flakes plays a key role in bringing about realistic devices, which has not been reported yet. In this work, we study the ferromagnetic properties of atomically thin $CrSiTe_3$ flakes using in-situ magnetic circular dichroism (MCD) microscopy and polarized Raman spectroscopy. We found that $CrSiTe_3$ experiences a magnetic transition from soft to hard ferromagnetic states as the thickness decreases down to several nanometers. It exhibits a rectangular hysteresis loop with finite magnetization and coercive field, indicating a single-domain, out-of-plane ferromagnetic order. The effective remanent magnetization in atomically thin $CrSiTe_3$ makes it possible to design ultrahigh-density data storage and nanoscale magnetic functional devices.

**Experimental methods**

$CrSiTe_3$ single crystals studied here are grown by the flux method (see supporting information (SI) for details). Fig. S1 shows a single-crystal x-ray diffraction (XRD) pattern with the presence of only a set of sharp peaks, indicative of high-quality samples. Magnetic susceptibility ($\chi$) measurements in $CrSiTe_3$ single crystal (Fig. 1b) reveal a paramagnetic-to-ferromagnetic phase transition with a Curie temperature of 32.8 K and an easy axis along the c-axis. In this work, atomically thin $CrSiTe_3$ flakes were prepared by mechanical exfoliation from bulk crystals. Fig. S2 shows the optical images and the atomic force microscope (AFM) images of exfoliated $CrSiTe_3$ samples with different thicknesses on $SiO_2$/Si substances. The thickness of the thinnest sample is confirmed to be around 1.7 nm

(around 2 layers). Since CrSiTe$_3$ belongs to air-sensitive materials[19], each step of sample preparation was finished in a glove box filled with nitrogen. In addition, to resolve the problem of sample degradation in the process of mounting the samples within the cryostat, we designed a sealed sample holder for the sample transfer and low-temperature measurements (see SI for details). The thickness-dependent ferromagnetic measurements in CrSiTe$_3$ were carried out using homemade MCD spectroscopy (see Methods for details) in reflection geometry. The applied magnetic field is along the c-axis of the sample with a resolution of 0.001 T. In contrast to traditional magneto-transport measurement, this method can avoid any degradation in the device fabrication process.

## Results and discussion

We first characterize the magnetic properties of bulk CrSiTe$_3$ using MCD spectroscopy. Fig. 1c shows the temperature-dependent MCD results in a 170 nm CrSiTe$_3$ flake. At 8 K, when the magnetic field ($H_\perp$) sweeps from 0.3 to -0.3 T, the MCD signal suddenly jumps a bit from saturation magnetization at around 0.16 T (the black arrow in Fig. 1c), and then changes linearly with the magnetic field and passes through zero point at zero magnetic field, and finally saturates above a negative magnetic field of -0.18 T, and vice versa. This special hysteresis loop has also been reported in another layered ferromagnet Fe$_3$GeTe$_2$[20], which is defined as an intermediate magnetic sate between single-domain ferromagnetism and paramagnetism. It is mainly induced by the formation of labyrinthine domains. This typical magnetic response in 2D materials has much common with that in traditional ferromagnetic films[21, 22]. The magnetic force microscopy (MFM) measurement in ref. [23] further confirms that labyrinthine-domain structures also form in bulk CrSiTe$_3$ below Curie temperature[23]. Here, when the $H_\perp$ decreases from the saturation field ($H_s$), labyrinthine-like nuclei start to generate, and further reduction of $H_\perp$ causes the nuclei to grow and spread, and saturate finally at the opposite $H_\perp$. Moreover, there is no measurable hysteresis loop, namely, remanent magnetization at $\mu_0 H_\perp = 0\,T$. It indicates the soft ferromagnetic property of bulk CrSiTe$_3$. Above ~32 K, MCD signals show a linear response between magnetization and applied field within ±0.3 T, revealing a soft-ferromagnetic to a paramagnetic state transition. The corresponding transition temperature is evaluated to be around 33±1 K, which is consistent with the results of susceptibility measurement in Fig. 1b.

Fig. 2a shows MCD signals as a function of applied field in atomically thin CrSiTe$_3$ flakes down to 1.7 nm at a fixed temperature of 8 K. As the thickness decreases to several tens of nanometers (29, 14, and 10.6 nm in Fig. 2a), the MCD signals still show a linearly magnetic response to the applied field with a saturation magnetization, similar with bulk one. However, in contrast to the closed hysteresis loop in the bulk one, a hysteresis loop gradually appears in these thin samples, which is characteristic of a wasp-waisted shape with a weak remanent magnetization. Importantly, the $H_s$ shows an obvious thickness dependence, gradually decreasing from 0.18 T in bulk to 0.04 T in 10.6 nm sample. The thickness dependence of $H_s$ for CrSiTe$_3$ flakes is summarized in Fig. 2b, where the threshold thickness for the reduction of $H_s$ is around 60 nm.

Surprisingly, as the thickness further reduces, a typical rectangular hysteresis loop with finite magnetization and coercive field appears in the samples at several nanometers (8 and 7 nm samples in Fig. 2a). It indicates that a single-domain, highly anisotropic out-of-plane ferromagnetic ordering forms. However, the $H_s$ is one order of magnitude lower than that in the bulk one. The average coercive field is evaluated to be 0.009 T for 8 nm and 0.0067 T for 7 nm samples at 8 K. A spatial mapping of the MCD intensity at $\pm 0.3\ T$ (Fig. S3) also reveals the magnetization to be near uniform across the entire flake. For the thinner samples, the hysteresis loop gradually narrows for the 5 nm sample and becomes negligible for the 4.2 nm sample. As thickness decreases below 3 nm (2.5 and 1.7 nm samples), MCD signals show a linear relationship with the applied field, indicating the absence of any long-range ferromagnetic orders at 8 K.

For the change from soft to hard ferromagnetism with decreasing thickness, a possible explanation is the suppression of the formation of the domain in thinner CrSiTe$_3$ samples. Theoretically, the thickness and energy of magnetic domain walls are two important parameters to decide the magnetic properties in two-dimensional materials. For the samples in the nanometer range of thickness, it is too thin to support a domain wall[24]. In CrSiTe$_3$, the size of the domain wall can be simply evaluated by the equation of $d = \pi\sqrt{A/K_u}$, where $A = JS^2/a$ is called the exchange stiffness with dimension. $J$, $S$ and $a$ are exchange interaction, spin moment, and lattice parameter, respectively. $K_u$ is magnetocrystalline anisotropy. Based on the DFT calculation[17], the domain wall thickness is evaluated to be around 10 nm for $J = 5\ meV$ and $K_u = 0.2\ meV/Cr$, respectively. Therefore, we suggest that it is difficult to form the domain wall in the samples below that value. It is consistent with the presence and absence of the maze-shaped domain in bulk and thin samples in our experimental observation, respectively. Moreover, from the energy point of view, ferromagnetic domains arise from the minimization of the stray field energy. In the bulk

CrSiTe$_3$, since the stray field energy is larger than the energy it takes to form domain walls, the ferromagnet will break up into domains. However, as the thickness decreases, the saturation field decrease as shown in Fig. 2a, as well as the magnetization and stray field. When the wall energy is larger than the stray field energy, small magnetic particles cannot support a domain wall, therefore such nanomagnets have to be a single domain, just as what we observed in 7 and 8 nm samples. The schematic diagram is shown in Fig. 2c. As to the disappeared ferromagnetism in samples 2.5 and 1.7 nm samples, we will discuss it later.

Fig. 3 shows the temperature dependence of MCD signals in the four representative samples, e.g., 29, 7, 5, and 4.2 nm thicknesses. As temperature increases, the ferromagnetic state with saturation magnetization gradually disappears, replaced by a linear paramagnetic response at high temperature. From that, we can simply evaluate the transition temperature ($T_c$) for all the samples as indicated by the red arrows in Fig. 3a and Fig. S4. Fig. 3b summarizes the thickness-dependent $T_c$ with error bars. It remains at around 33±1 K for the samples thicker than 8 nm. This similar temperature dependence suggests they share a similar ferromagnetic origin. Below that, it dramatically decreases, where the $T_c$ for 5 and 4.2 nm samples drops down to around 26.5±1.5 K and 17±3 K, respectively. This thickness-dependent behavior indicates a crossover from 3D to 2D Ising ferromagnetism, which is often observed in ultrathin layered magnets, such as 2D magnets Fe$_3$GeTe$_2$ [20], CoPS$_3$ [25].

Another possibility for the layer-dependent magnetic properties is the structure transition or distortion as thickness decreases, as reported in the literature [10]. This can be checked by using thickness-dependent Raman measurements (Fig. S5). Fig. 4a shows the linearly polarized Raman spectra in a 14 nm CrSiTe$_3$ flake at 8 K, where parallel (XX) and perpendicular (XY) configurations are indicated by red and black curves, respectively. Seven unambiguous peaks (indicated by dashed line) are identified with Raman modes labeled in Fig. 4a. Here, we assign three modes at 82, 149, and 518 cm$^{-1}$ to non-degenerate A$_g$ vibration modes and four peaks at 89, 119, 218, and 363 cm$^{-1}$ to double-degenerate E$_g$ modes. Our Raman results are consistent with the results reported in literature [26-29]. According to factor-group analysis for the $R\bar{3}$ space group[28], there are five A$_g$ modes and five doubly degenerate E$_g$ modes. All the Raman modes are active and should be observed with excited light propagating along the c-axis. The absence of peaks at 96.9 ($E_g^2$), 122 ($A_g^2$), and 208.7 ($A_g^4$) may be due to weak scattering cross-section or non-resonant excitation. Fig. S6 shows the temperature-dependent Raman spectra in the 14 nm sample. As temperature increases, all the Raman peaks exhibit a slight redshift, but there is no sudden change near Curie temperature. It indicates that any structural phase transition is absent during paramagnetic-to-ferromagnetic phase transition. Moreover, the polarization-dependent Raman measurement in 14 nm $CrSiTe_3$ at 8 K (Fig. S7) shows a polarization-angle independent response, indicating the absence of any structure distortion at the ferromagnetic states.

Fig. 4b shows the Raman curves in several fixed thicknesses at 8 K. With decreasing thickness, six low energy peaks still can be measurable with unchanged peak energy in samples down to 4 nm. It indicates that the crystal structure remains unchanged, hence, we can also exclude any structural transition for thin-layer samples. However, for the 2.5 nm sample, the intensity of all the Raman peaks decreases. It is very difficult to identify any sharp Raman peak for 1.7 nm samples. It may be induced by semiconductor-to-metal transition or structure instability[19] for fewer layer samples, which is consistent with disappearing ferromagnetic ordering in the MCD measurements. Moreover, some calculations indicate that monolayer $CrSiTe_3$ will transfer to the antiferromagnetic state with a zigzag spin texture [30], which cannot be measured using MCD spectroscopy. It notes that an extra peak at ~125 $cm^{-1}$ appears in 7 and 8 nm samples, as indicated by the dashed square in Fig. 4b. To trace the cause, we conducted Raman measurement in the 7 nm sample after exposing it to the air at different times. As shown in Fig. 4c, the peak at ~125 $cm^{-1}$ gradually enhances for ~30 min exposure, indicating its origin from the $TeO_2$ due to oxidation of Te atom in the air [19, 28]. However, the MCD measurements (Fig. S8) still reveal a rectangular hysteresis loop at low temperature desperate exposure to the atmosphere. The evaluated $T_c$ remains unchanged in comparison with the sample without exposure. It suggests that the ferromagnetism of $CrSiTe_3$ flakes is very robust after short-term exposure because of its origin from the centered Cr atom.

## Summary

In conclusion, we have observed the behavior of soft ferromagnetic to hard ferromagnetic transition with decreasing thickness of $CrSiTe_3$ flakes. A rectangular hysteretic loop, signed by a remanent magnetization at zero field, appears in the 7 and 8 nm samples at 8 K. It indicates a single domain and out-of-plane ferromagnetic order. We suggest the suppression of domain walls contributes to this phenomenon. Our results will help to understand the underlying mechanism of magnetic exchange interaction in 2D materials.

**Acknowledgment**

We would like to thank Prof. Haizhou Lu from SUSTech for helpful discussions. J.F.D. acknowledges the support from the National Natural Science Foundation of China (11974159) and the Guangdong Natural Science Foundation (2021A1515012316). J.W.M was partially supported by the program for Guangdong Introducing Innovative and Entrepreneurial Teams (No. 2017ZT07C062), and Shenzhen Key Laboratory of Advanced Quantum Functional Materials and Devices (No. ZDSYS20190902092905285). J.W.M and L.W. were supported by Guangdong Basic and Applied Basic Research Foundation (No. 2020B1515120100). L.W. was supported by China Postdoctoral Science Foundation (2020M682780).

**Conflict of Interest**

The authors declare no conflict of interest.

**Keywords**

2D ferromagnetic materials, Raman spectroscopy, magnetic circular dichroism (MCD) spectroscopy

**Methods**

**Magnetic circular dichroism measurements**

The ferromagnetic behaviors of the $CrSiTe_3$ flake were measured in magnetic circular dichroism (MCD) system. Here, a 632.8 nm HeNe laser was employed as an excited light, which is focused onto the sample using a 50x objective with a spot size of around 2 μm. A photoelastic modulator (PEM) was set on the path of the laser beam, so that the polarized state of excitation can change between left-handed and right-handed circularly polarization with a frequency of 50 kHz. The reflected light from the sample surface was fed to a silicon detector and read out by a lock-in amplifier, which trach the frequency of PEM. Therefore, the intensity difference between two circularly polarized lights can be recorded, which is proportional to the magnetization of materials. To get magnetism-related signals, an out-of-plane magnetic field, generated from a superconducting loop, was swept within $\pm 0.3$ T. Hence, we can get the MCD intensity as a function of the applied magnetic field to identify the ferromagnetic properties of materials.

**Polarized Raman spectroscopy**

The schematic diagram of polarized Raman measurement is shown in Fig. S5. A 532 nm solid-state laser was used as an excitation. The laser beam passed through a polarizer, a beam splitter, and a 50x objective, and then focused onto the sample with a spot size of around 2 $\mu m$. The laser power on the sample is around 200 $\mu W$. The back-scattered Raman signal was collected by the same objective, passed through the beam splitter and a 532 nm edge filter, and then was focused on the entrance slit of a spectrometer equipped with a nitrogen-cooled charge-coupled device (CCD). To conduct the polarization-dependent measurements, another polarizer was fixed before the spectrometer. By rotating the polarized direction of this polarizer, we can get the Raman signals for both parallel- and crossed-polarization configurations. For the polarized Raman measurement, two polarizers were kept at parallel-polarization configuration and a half-wave plate was located between the beam splitter and objective. By rotating the half-wave plate, we can change the polarization angle between the x-axis of the laboratory coordinate and the polarization direction of the laser beam. Hence, the polarized Raman modality can be obtained and employed to characterize the crystal structure of 2D van der Waals materials.

**FIGURES AND CAPTIONS**

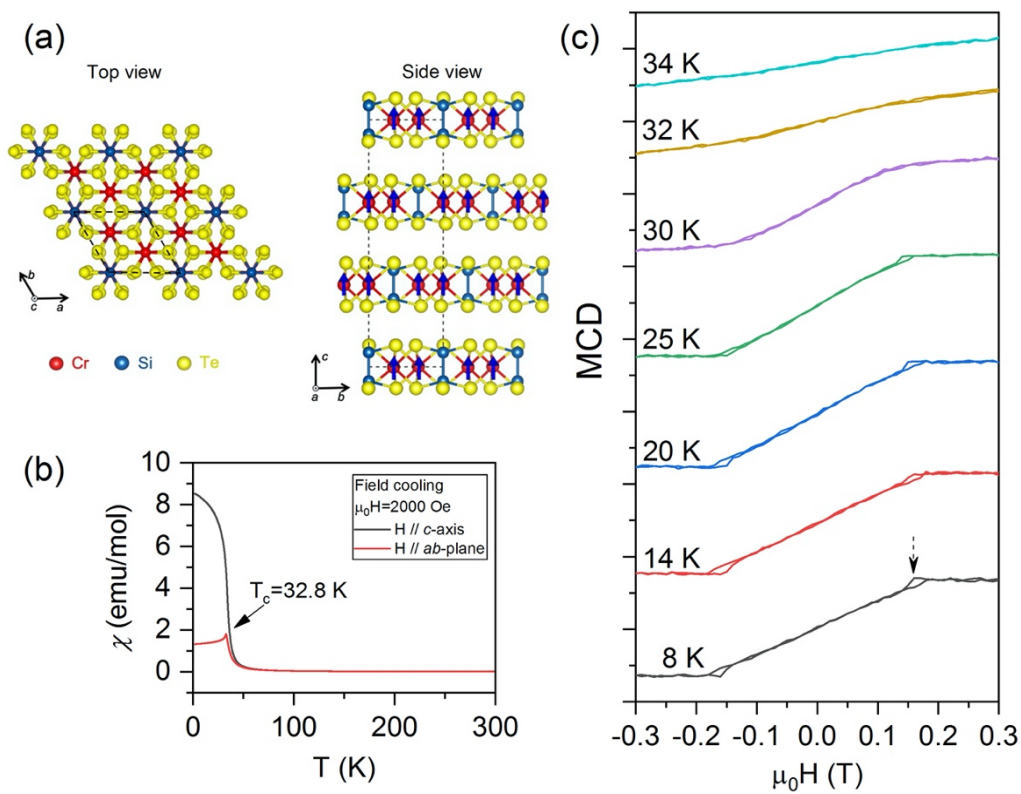

Figure 1: (a) Top and side views of the crystal structure of $CrSiTe_3$. Blue arrows indicate the spin orientation in Cr atoms. (b) Magnetic susceptibility of $CrSiTe_3$ measured on a single crystal at the field cooling (FC) along the c-axis (black curve) and ab-plane (red curve), respectively. (c) MCD signals as a function of applied magnetic field ($H_\perp$) along c-axis in a 170 nm $CrSiTe_3$ flake at several fixed temperatures. The black arrow indicates the magnetic field for the jump of saturation magnetization.

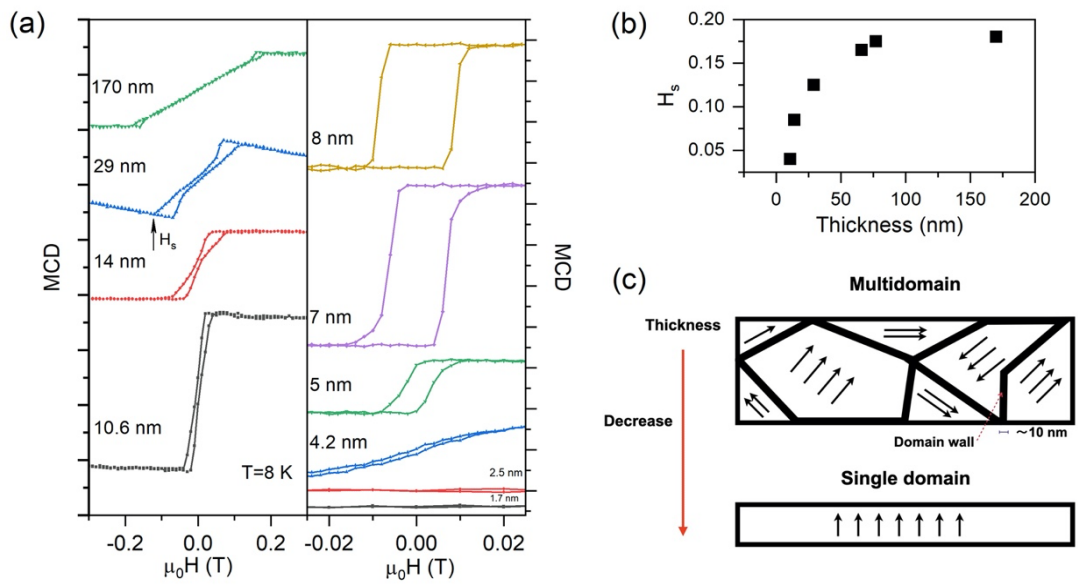

Figure 2: (a) MCD signals as a function of the applied magnetic field in CrSiTe$_3$ flakes with different thicknesses. The black arrow indicates the saturation field (H$_s$). (b) Extracted saturation field as a function of thickness. (c) Schematic diagram of the evolution of domain as thickness decreases.

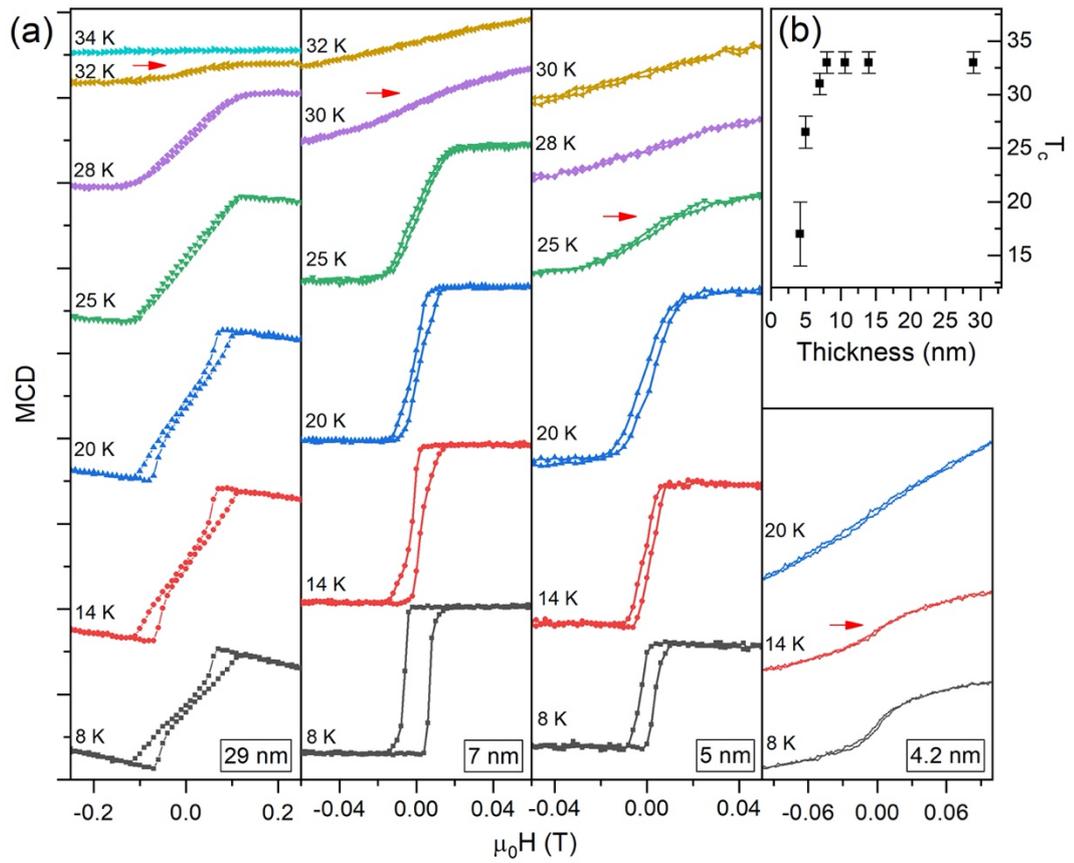

Figure 3: (a) Temperature dependence of MCD signals in four representative thicknesses of CrSiTe$_3$ flakes, e.g., 29 nm, 7 nm, 5 nm, and 4.2 nm. The red arrows indicate the temperature for phase transition from ferromagnetic to paramagnetic states. (b) Thickness dependence of $T_c$ in CrSiTe$_3$.

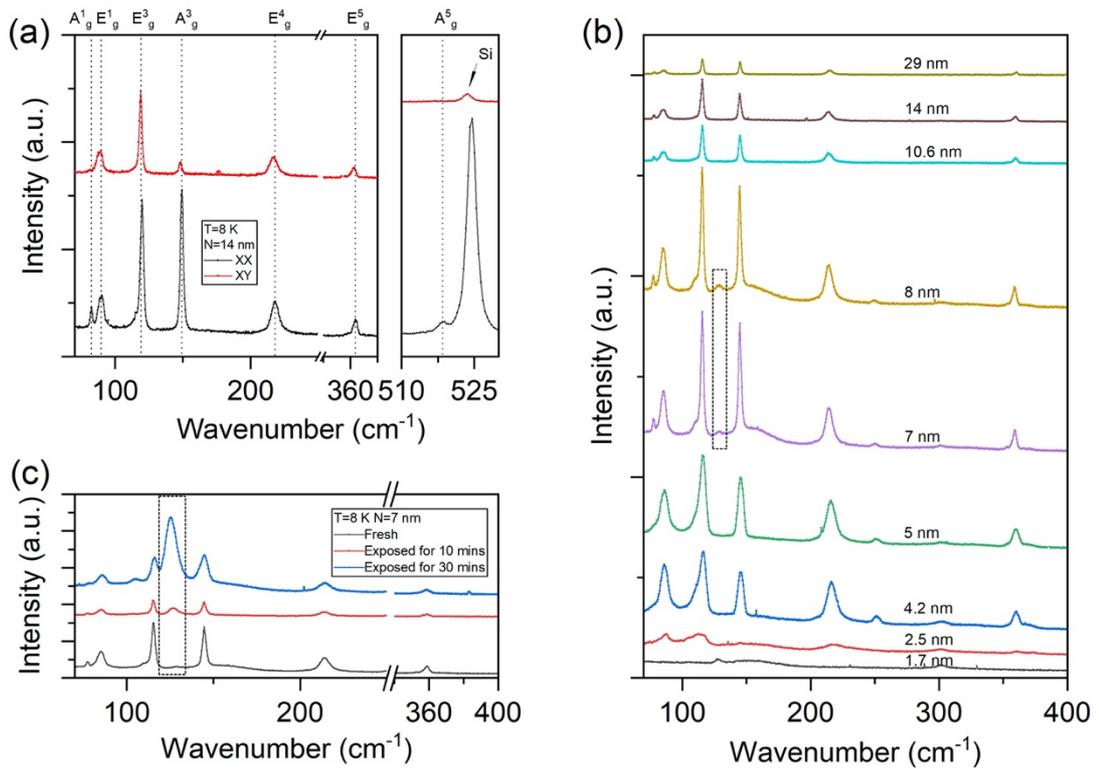

Figure 4: (a) linearly polarized Raman spectra of a 14 nm CrSiTe$_3$ measured at 8 K in parallel (XX) and cross (XY) polarization scattering configurations, respectively. The dashed lines indicate the position of peaks and the corresponding Ramon modes are also labeled. The peak at 525 cm$^{-1}$ belongs to the characteristic peak of Si substrate. (b) Raman spectra in CrSiTe$_3$ flakes with different thicknesses. The broad peak at around 300 cm$^{-1}$ comes from the Si substrate. (c) Raman spectra in the 7 nm CrSiTe$_3$ before and after exposure to the air. The peaks in the dashed square come from the TeO$_2$.

Cobden, D. H.; Dean, C. R.; Xiao, D.; Xu, X., Switching 2D magnetic states via pressure tuning of layer stacking. *Nature Materials* **2019,** *18* (12), 1298-1302.

10. Sun, Z.; Yi, Y.; Song, T.; Clark, G.; Huang, B.; Shan, Y.; Wu, S.; Huang, D.; Gao, C.; Chen, Z.; McGuire, M.; Cao, T.; Xiao, D.; Liu, W.-T.; Yao, W.; Xu, X.; Wu, S., Giant nonreciprocal second-harmonic generation from antiferromagnetic bilayer CrI3. *Nature* **2019,** *572* (7770), 497-501.

11. Kim, K.; Lim, S. Y.; Lee, J.-U.; Lee, S.; Kim, T. Y.; Park, K.; Jeon, G. S.; Park, C.-H.; Park, J.-G.; Cheong, H., Suppression of magnetic ordering in XXZ-type antiferromagnetic monolayer NiPS3. *Nature Communications* **2019,** *10* (1), 345.

12. Chu, H.; Roh, C. J.; Island, J. O.; Li, C.; Lee, S.; Chen, J.; Park, J.-G.; Young, A. F.; Lee, J. S.; Hsieh, D., Linear Magnetoelectric Phase in Ultrathin ${\mathrm{MnPS}}_{3}$ Probed by Optical Second Harmonic Generation. *Physical Review Letters* **2020,** *124* (2), 027601.

13. Williams, T. J.; Aczel, A. A.; Lumsden, M. D.; Nagler, S. E.; Stone, M. B.; Yan, J. Q.; Mandrus, D., Magnetic correlations in the quasi-two-dimensional semiconducting ferromagnet ${\text{CrSiTe}}_{3}$. *Physical Review B* **2015,** *92* (14), 144404.

14. Zhang, J.; Cai, X.; Xia, W.; Liang, A.; Huang, J.; Wang, C.; Yang, L.; Yuan, H.; Chen, Y.; Zhang, S.; Guo, Y.; Liu, Z.; Li, G., Unveiling Electronic Correlation and the Ferromagnetic Superexchange Mechanism in the van der Waals Crystal ${\mathrm{CrSiTe}}_{3}$. *Physical Review Letters* **2019,** *123* (4), 047203.

15. Ron, A.; Zoghlin, E.; Balents, L.; Wilson, S. D.; Hsieh, D., Dimensional crossover in a layered ferromagnet detected by spin correlation driven distortions. *Nature Communications* **2019,** *10* (1), 1654.

16. Carteaux, V.; Moussa, F.; Spiesser, M., 2D Ising-Like Ferromagnetic Behaviour for the Lamellar Cr 2 Si 2 Te 6 Compound: A Neutron Scattering Investigation. *Europhysics Letters (EPL)* **1995,** *29* (3), 251-256.

17. Zhang, C.; Gu, Y.; Wang, L.; Huang, L.-L.; Fu, Y.; Liu, C.; Wang, S.; Su, H.; Mei, J.-W.; Zou, X.; Dai, J.-F., Pressure-Enhanced Ferromagnetism in Layered CrSiTe3 Flakes. *Nano Letters* **2021,** *21* (19), 7946-7952.

18. Dilip, B. J., Gouchi; Naoka, Hiraoka; Yufeng, Zhang; Norio, Ogita; Takumi, Hasegawa; Kentaro, Kitagawa; Hidenori, Takagi; Kee Hoon, Kim; Yoshiya, Uwatoko, Nearly room temperature ferromagnetism in pressure-induced correlated metallic state of van der Waals insulator CrGeTe$_3$. *arXiv* **2021,** *2107.10573*.